\documentclass[11pt]{article}
\usepackage[T1]{fontenc}

\providecommand{\keywords}[1]{\textbf{\textit{Keywords:---}} #1}

\usepackage{amsfonts}
\usepackage{amsthm}
\usepackage{comment}
\usepackage{amsmath}
\usepackage{amssymb}
\usepackage{cases}
\usepackage{mathtools}
\usepackage{graphicx}

\usepackage{plain}

\def\pe{p^\varepsilon}
\def\qe{q^\varepsilon}

\newcommand{\dis}{\displaystyle}

\newcommand{\mmmintone}[1]{{\dis{\int\kern -.36cm-}}_{\kern-.21cm\substack{#1}}\;\;}
\newcommand{\mmmintwo}[2]{{\dis{\int\kern -.43cm-}}_{\kern-.21cm\substack{#1}}^{\substack{#2}}\;\;}
\newcommand{\submint}{{\scriptstyle{\int\kern -.66em -}}}
\newcommand{\submintone}[1]{{\scriptstyle{\int\kern -.66em-}}_{\scriptscriptstyle{\kern-.21em\substack{#1}}}}
\newcommand{\fracmint}{{\textstyle{\int\kern -.88em -}}}
\newcommand{\fracmintone}[1]{{\textstyle{\int\kern -.88em
-}}_{\scriptscriptstyle{\kern-.21em\substack{#1}}}\;}

\title{Dissipation functions and Brownian oscillators}

\author{M. Colangeli \footnote{Dipartimento di Ingegneria e Scienze dell'Informazione e Matematica, Universit\`{a} degli Studi dell’Aquila, Via Vetoio, 67100 L’Aquila, Italy; e-mail: matteo.colangeli1@univaq.it}, L. Rondoni\footnote{Dipartimento di Scienze Matematiche, Politecnico di Torino, Corso Duca degli Abruzzi 24, 10129 Torino,  Italy; e-mail: lamberto.rondoni@polito.it}\footnote{INFN, Sezione di Torino, Via Pietro Giuria 1, 10125 Torino, Italy.}, P. Vozza\footnote{Universit\`{a} degli Studi di Torino, Dipartimento di Fisica, Via Pietro Giuria 1, 10125 Torino, Italy; e-mail: pasquale.vozza@edu.unito.it}}

\date{\today}

\begin{document}

\maketitle

\begin{abstract}
\noindent
We develop a general framework for response theory in diffusion processes governed by Fokker-Planck equations, based on the notion of the Dissipation Function. Using the analytically solvable Brownian oscillator model, we derive exact response formulae for both overdamped and underdamped dynamics of harmonically bound Brownian particles. We also demonstrate that for certain observables and under suitable time scaling, the operations of model reduction and response formula extraction commute.
\end{abstract}

\noindent
\keywords{Response theory; Fokker-Planck equation; overdamped limits; Underdamped Langevin dynamics.} 

\section{Introduction}

Since the seminal contributions of  Kubo~\cite{Kubo66,Kubo}, Linear Response Theory has become a foundational framework in nonequilibrium statistical mechanics. Initially formulated to address Hamiltonian perturbations in deterministic systems~\cite{Zwanzig}, subsequent developments have significantly expanded its scope to incorporate nonlinear corrections~\cite{Bochkov1981,LucCol12}, as well as non-Hamiltonian dynamics and dissipative stochastic processes~\cite{Ruelle1998,Ruelle1999,Vulp,Prost2009,Baiesi2009,Maes09,Maes10,Seifert2010,colan11,colan12,Maes20,MaesPR2020,Baldovin2022}; see also \cite{Nyquist1928,Johnson1928,Callen1951} for a historical perspective and \cite{Bellon2001,Wang2002,Ojha2004,Feitosa2004,Carberry2004,Greinert2006,Mizuno2007,GomezSolano2009,Ciliberto2017} for experimental validations. 
The search for non-perturbative approaches to response theory, such as the method based on the transient time correlation function developed in~\cite{EM90}, gained renewed momentum with the introduction of the \emph{Dissipation Function}, defined in~\cite{ES00,ES02} as the energy dissipation rate that satisfies the fluctuation relation.
 This concept enabled the unification, within a consistent theoretical framework, of the diverse methodologies employed in nonequilibrium molecular dynamics. The Fluctuation Theorem, systematized in~\cite{ESW08}, provides a rigorous probabilistic description of entropy production in small, driven systems and applies to time-reversible dynamics subjected to constant external forces. Extensions to periodically driven systems were subsequently developed in~\cite{PE97a,PE97b,Todd98}. The mathematical framework of response theory based on the Dissipation Function was further formalized in~\cite{Jepps16}, and later extended to quantum dynamics~\cite{Rond25}, as well as to systems exhibiting collective behavior and phase transitions, such as Kuramoto-type models~\cite{ACCR22}, where standard linear response theory typically breaks down.

In this work, we extend this formalism to stochastic diffusion processes with additive noise. We demonstrate that many structural features originally derived for deterministic dynamics can be systematically transferred and adapted to stochastic systems governed by Fokker--Planck equations \cite{VK07,Gardiner09}.

We further derive explicit response formulae for both overdamped and underdamped Langevin dynamics. Importantly, these formulae are \emph{exact}, in the sense that they are not limited to the linear response regime and remain valid for finite perturbation amplitudes.

Focusing on the analytically tractable case of harmonic Brownian oscillators, we illustrate the applicability of the exact response theory, that becomes particularly interesting in complex situations. We then examine the commutativity of two operations: the derivation of response expressions and the overdamped limit, which corresponds to a perfect separation of time scales between fast and slow variables. This commutativity for selected observables reflects a structural symmetry of the response formalism based on the Dissipation Function and may also have practical implications. In particular, for systems with complex particle dynamics, it may be more tractable to probe the regime of perfect time scale separation at the level of observable averages, rather than by resolving the full stochastic dynamics of individual constituents.

The paper is organized as follows. In Sec.~\ref{sec:sec2}, we introduce the response theory based on the Dissipation Function. Section~\ref{sec:sec3} presents the formal derivation of the overdamped limit of underdamped Langevin dynamics. In Sec.~\ref{sec:sec4}, we specialize to the harmonic case and derive exact response formulae for the overdamped dynamics of Brownian oscillators subjected to a perturbation that simultaneously alters the frequency of the external field and the temperature of the surrounding thermal bath. Section~\ref{sec:sec5} addresses the underdamped case and analyzes the limiting behavior of the response in the regime of infinite time scale separation. Conclusions are drawn in Sec.~\ref{sec:sec6}, while technical details are provided in the Appendices.

\section{Dissipation Function in the Fokker–Planck Equation}
\label{sec:sec2}
In this section, we analyze the role of the Dissipation Function in the evolution of probability densities governed by the Fokker–Planck equation.
We begin with the stochastic differential equation (SDE) for a variable \( x \in \mathbb{R}^d \) written in the It\^o's form:
\begin{equation}
    dx(t) = A(x)\, dt + \sigma(x)\, dW(t) \, , \quad x(0)=x_0  , \label{pertSDE}
\end{equation}
where \( A: \mathbb{R}^d \to \mathbb{R}^d \), \( \sigma: \mathbb{R}^d \to \mathbb{R}^{d \times m} \), and \( W(t) \) is a standard Wiener process in \( \mathbb{R}^m \).
The generator \( L \) of the process acts on state functions \( O: \mathbb{R}^d \to \mathbb{R} \) as follows:
\begin{equation}
    L\ O(x) = A(x) \cdot \nabla O(x) + D(x) : \left( \nabla \nabla O(x) \right)  , \label{gen}
\end{equation}
where \( D(x) = \frac{1}{2} \sigma(x) \sigma(x)^T \) is the diffusion matrix, and \( \nabla \nabla O \) denotes the Hessian matrix of \( O \).

The \( L^2 \)-adjoint of the generator \( L \), known as the \textit{Fokker–Planck operator}, acts on probability densities \( f(x,t) \) according to:
\begin{equation}
    L^* f(x,t) = \nabla \cdot \left[ -A(x) p(x,t) + \nabla \cdot \left( D(x) f(x,t) \right) \right] \, .
    \label{adj}
\end{equation}
The corresponding forward Kolmogorov equation (or Fokker–Planck equation) for \( f(x,t) \) reads:
\begin{subequations}
\label{FP}
\begin{align}
\partial_t f(x,t) &= L^* f(x,t) \;\, , \quad \text{in}\ \mathbb{R}^d \times (0,\infty)  , \label{FP1} \\
f(x,0) &= f_0(x) \qquad , \quad \text{on}\ \mathbb{R}^d\times\{t=0\}  .
\end{align}
\end{subequations}
Using the shorthand notation \( f_t(x) = f(x,t) \), Eq.~\eqref{FP1} can be rewritten as:
\begin{equation}
    \partial_t f_t(x) = \Omega^{f_t}(x)\, f_t(x) \, , \label{forwop}
\end{equation}
where the state function \( \Omega^{f_t}(x) := (L^* f_t(x))/f_t(x) \), called \textit{Dissipation Function}\footnote{We purposely adopt, here, the same terminology used in the deterministic setting \cite{Jepps16}.}, is given explicitly by:
\begin{align}
\Omega^{f_t}(x) 
&= -\nabla \cdot A(x) 
- A(x) \cdot \nabla \log f_t(x) 
+ \nabla \nabla : D(x) \notag \\
&\quad + 2\, (\nabla \cdot D(x)) \cdot \nabla \log f_t(x) \notag \\
&\quad + D(x) : \left[ \nabla \nabla \log f_t(x) + \nabla \log f_t(x) \otimes \nabla \log f_t(x) \right] ,
\label{Om}
\end{align}
see Appendix~\ref{app:appA} for details.
Finally, we define the expectation of the state function \( O(x) \) at time \( s \geq 0 \) as:
\begin{equation}
    \langle O \rangle_s = \int_{\mathbb{R}^d} O(x)\, f_s(x)\, dx  . \label{aver}
\end{equation}
Exploiting the formal solution of Eq. \eqref{FP1}, we find:
\[
\langle O \rangle_s = \int O(x) \left(e^{s L^* }p_0(x)\right) \, dx = \int \left(e^{s L} O(x)\right) p_0(x) \, dx.
\]
The time derivative of $\langle O \rangle_s$ reads:
\[
\frac{d}{ds} \langle O \rangle_s = \int \left( \frac{d}{ds} e^{s L} O(x) \right) f_0(x) \, dx = \int (L e^{s L} O(x)) f_0(x) \, dx,
\]
which, using the adjoint operator, becomes:
\[
\frac{d}{ds} \langle O \rangle_s = \int \left(e^{s L} O(x)\right) \left(L^* f_0(x)\right) \, dx .
\]
Finally, from \eqref{forwop}, we find:
\begin{equation}
    \frac{d}{ds} \langle O \rangle_s 
= \int \left(e^{s L} O(x)\right) \Omega^{f_0}(x) f_0(x) \, dx = \langle O(x_s) \Omega^{f_0}(x_0) \rangle_0 .
\label{tder}
\end{equation}
By integrating both sides of \eqref{tder} from 0 to \( t \), we finally obtain the Exact Response Formula:
\begin{equation}
    \langle O \rangle_t - \langle O \rangle_0 = \int_0^t \langle O(x_s) \Omega^{f_0}(x_0)  \rangle_0 \, ds . \label{final}
\end{equation}
\noindent
\noindent
We note that the average appearing under the integral in Eq.~\eqref{final} is taken with respect to the initial density \( f_0(x) \). To establish a connection with response theory, it is helpful to further characterize \( f_0(x) \) introducing an alternative process, called the \textit{reference dynamics} and defined for times \( t \in (-\infty, 0) \), which admits \( f_0(x) \) as its invariant probability density. The dynamics governed by Eq.~\eqref{pertSDE} can be interpreted as the result of applying a perturbation to the reference dynamics at time \( t = 0 \), and therefore will be referred to as the \textit{perturbed dynamics}.

We also observe that if in Eq. \eqref{final} one sets $O(x)=1$ for all $x\in \mathbb{R}^d$, then the normalization of the probability density $f_s(x)$, for $s\ge 0$, implies $\langle \Omega^{f_0} \rangle_0 =0$.




\section{From the underdamped to the overdamped Langevin dynamics}
\label{sec:sec3}

We let \( x = (q, p)^T \in \mathbb{R}^2 \) and consider the underdamped Langevin equation describing a Brownian particle subjected to an external potential \( U(q) \) \cite{Risken,Pavl}:
\begin{subequations}
\label{eq:LE0}
\begin{align}
d q(t) &= \frac{p}{m}\, dt , \label{eq:LE0-1} \\
d p(t) &= - \nabla_{q}U(q) \, d t - \gamma~ p\, dt + \sqrt{2\gamma m \beta^{-1}}\, d W(t) . \label{eq:LE0-2}
\end{align}
\end{subequations}
We briefly outline the formal steps leading to the overdamped limit of Eqs. \eqref{eq:LE0}, following the standard derivation described in \cite{Pavl,colan25}.
We introduce $\epsilon>0$ and in Eqs.~\eqref{eq:LE0} we rescale \( \gamma \mapsto \gamma/\varepsilon \) and \( t \mapsto t/\varepsilon \), which yields
\begin{subequations}
\label{eq:LEsc}
\begin{align}
d \qe &= \frac{\pe}{\varepsilon m}\, dt , \label{eq:LE-1} \\
d \pe &= - \frac{1}{\varepsilon} \nabla_{q}U(\qe)\, dt - \frac{\gamma}{\varepsilon^2} \pe\, dt + \frac{1}{\varepsilon} \sqrt{2\gamma m \beta^{-1}}\, d W(t) , \label{eq:LE-2}
\end{align}
\end{subequations}
where we have exploited the scaling property of the Wiener process \( W(t/\varepsilon) = 1/\sqrt{\varepsilon}\ W(t) \).
From \eqref{eq:LE-1} and \eqref{eq:LE-2}, we obtain:
\[
d\qe = \frac{\pe}{\varepsilon m}\, dt = -\frac{\varepsilon}{\gamma m} d\pe - \frac{1}{\gamma m} \nabla_q U(\qe)\, dt + \sqrt{\frac{2}{m\beta \gamma}}\, dW(t).
\]
Taking the limit \( \varepsilon \to 0 \), we arrive at the overdamped Langevin dynamics \cite{Pavl}
\begin{equation}
dq = -\frac{1}{\gamma m} \nabla_q U(q)\, dt + \sqrt{\frac{2}{m\beta \gamma }}\, dW(t) .
\label{eq:LE}
\end{equation}
In the sequel we will focus on Brownian oscillators, namely Brownian particles bound by a harmonic potential $U(q)=\frac{1}{2}m\omega^2 q^2$, and evaluate Eq.~\eqref{final} for both the overderdamped dynamics \eqref{eq:LE} (in Sec.~\ref{sec:sec4}) and its underdamped counterpart \eqref{eq:LE0} (in Sec.~\ref{sec:sec5}).

\section{Exact response for the overdamped Langevin dynamics}
\label{sec:sec4}

The application of the response theory introduced in Sec. \ref{sec:sec2} to Eq. \eqref{eq:LE} begins with the definition of the reference dynamics:
\begin{equation}
    dq(t) = -\frac{\omega_0^2 q}{\gamma} \, dt + \sqrt{\frac{2}{m\beta_0 \gamma }}\, dW(t)  , \quad t\le 0 ,\label{smol0}
\end{equation}
which switches, at time $t=0^+$, into the perturbed one, defined as follows:
\begin{equation}
    dq(t) = -\frac{\omega^2 q}{\gamma} \, dt + \sqrt{\frac{2}{m\beta \gamma }}\, dW(t) \; , \quad t>0  .\label{smol}
\end{equation}
Note that both parameters $\omega_0\rightarrow\omega$ and $\beta_0\rightarrow\beta$ have changed as a result of the perturbation.
The Fokker-Planck equation corresponding to Eq. \eqref{smol}, known as Smoluchowski equation \cite{Bala}, reads 
\begin{equation}
    \frac{\partial f_t(q)}{\partial t}=\frac{\partial}{\partial q}\left[ \frac{\omega^2}{\gamma} q\ f_t(q)+\frac{1}{m\beta \gamma}\frac{\partial}{\partial q}f_t(q) \right] ,
    \label{FP2}
\end{equation}
and is supplied with the initial datum 
\begin{equation}
f_0(q)=\sqrt{\frac{m \beta_0\omega_0^2}{2 \pi }}\exp{\left(-\frac{1}{2}m\beta_0\omega_0^2 q^2\right)} .
\label{f0}
\end{equation}
Equation \eqref{FP2} can be cast in the form
\begin{equation}
  \frac{\partial f_t(q)}{\partial t}=\Omega_{\omega,\beta}^{f_t} f_t , 
\end{equation}
where the subscripts in $\Omega_{\omega,\beta}^{f_t}$ highlight the two parameters that characterize the perturbation.
A straightforward computation (see Eq. \eqref{Om5} in Appendix \ref{app:appA}) leads to the following expression of the Dissipation Function $\Omega_{\omega,\beta}^{f_0}$ characterizing the overdamped dynamics \eqref{smol}:
\begin{equation}
\Omega_{\omega,\beta}^{f_0}(q)=A_0+B_0 q^2 \; ,
\label{Omover}
\end{equation}
where $A_0=(\beta\omega^2-\beta_0 \omega_0^2)/(\beta \gamma)$ and $B_0=-A_0/\langle q^2\rangle_0$, with $\langle q^2\rangle_0=(\beta_0 m\omega_0^2)^{-1}$.
Formula \eqref{Omover} reveals that the initial distribution \eqref{f0} is effectively invariant for the reference dynamics \eqref{smol0}, because it holds
\begin{equation}
 \Omega_{\omega_0,\beta_0}^{f_0}(q) = 0 .
\end{equation}
Moreover, we also point out that \( \Omega_{\omega,\beta}^{f_0} \) vanishes, more generally, if the condition \( \beta \omega^2 = \beta_0 \omega_0^2 \) is satisfied. This follows from the fact that the invariant densities of the reference and the perturbed dynamics are both Gaussian distributions:
\[
f_0(q) = \mathcal{N}\left(0,\frac{1}{\beta_0 m\omega_0^2}\right), \quad f_{\infty}(q) = \mathcal{N}\left(0,\frac{1}{\beta m \omega^2}\right).
\]
Under the condition \( \beta \omega^2 = \beta_0 \omega_0^2 \), the variances of these distributions coincide (and so do also the respective means, which both vanish), implying \( f_0(q) \equiv f_{\infty}(q) \).
For the Brownian oscillator model described by Eq. \eqref{smol}, the expression of the conditional probability density $p(r,t|q,0)$ can be made explicit:

\begin{equation}
    p(r,t|q,0)=\sqrt{\frac{\beta m\omega^2}{2 \pi \left(1-e^{-\frac{2\omega^2}{\gamma}t}\right)}}\exp{\left[-\frac{\beta m\omega^2 \left(r-q\,e^{-\frac{\omega^2}{\gamma}t} \right)^2}{2 \left(1-e^{-\frac{2\omega^2}{\gamma}t}\right)}\right]} .
    \label{condprob1}
\end{equation}
We compute, first, the exact response formula for the observable $O(q)\equiv q^2$.
In this case Eq. \eqref{final} reads
\begin{equation}
    \langle q^2 \rangle _t = \langle q^2 \rangle _0 + \int_0^t \langle q_s^2 \, \Omega_{\omega,\beta}^{f_0}(q_0)\rangle _0  \, ds  .
    \label{resex1}
\end{equation}
From Eq.~\eqref{condprob1}, we obtain the expression
\begin{equation}
 \langle q^2 \rangle _t= \frac{\beta \omega^2 e^{-2\kappa t}+\left(1-e^{-2\kappa t}\right)\beta_0 \omega_0^2}{m \beta_0 \beta \omega_0^2 \omega^2} \label{varover}  ,
\end{equation}
where $\kappa=\omega^2/\gamma$ denotes a characteristic relaxation frequency.
We point out that formula \eqref{varover} possesses the proper asymptotic behavior: it reduces to $\langle q^2 \rangle_0$ at $t=0$ while it recovers the expression $\langle q^2 \rangle_{\infty}= (\beta m\omega^2)^{-1}$ in the long time limit.
We turn, then, to the analysis of the response formula for the state function $O(q)\equiv\Omega_{\omega,\beta}^{f_0}(q)$.
In this case formula \eqref{final} reads:
\begin{equation}
    \langle \Omega_{\omega,\beta}^{f_0} \rangle _t = \langle \Omega_{\omega,\beta}^{f_0} \rangle _0 + \int_0^t \langle \Omega_{\omega,\beta}^{f_0}(q_s) \, \Omega_{\omega,\beta}^{f_0}(q_0)\rangle _0  \, ds .
    \label{resex2}
\end{equation}
We first verify that it correctly holds
\begin{equation}
  \langle \Omega_{\omega,\beta}^{f_0} \rangle _0=A_0+B_0\langle q^2\rangle_0=0 ,  
\end{equation}
whereas 
\begin{eqnarray}
   \int_0^t\langle \Omega_{\omega,\beta}^{f_0}(q_s) \,\Omega^{f_0}(q_0)\rangle _0  ds &=& B_0\int_0^s \langle q_s^2, \Omega_{\omega,\beta}^{f_0}(q_0) \rangle_0 ds \nonumber\\
   &=&B_0(1-e^{-2 \kappa t})(\langle q^2\rangle_{\infty}-\langle q^2\rangle_0) \nonumber\\
   &=& \frac{(\beta \omega^2-\beta_0\omega_0)^2}{\beta^2 \gamma\omega^2}(1-e^{-2 \kappa t})
   \label{resp} .
\end{eqnarray}


\section{Exact response theory for the underdamped Langevin dynamics}
\label{sec:sec5}

We now apply the formalism of Sec.~\ref{sec:sec2} in the set-up of the underdamped dynamics \eqref{eq:LE0}. We start, as in Sec. \ref{sec:sec4}, with the definition of the reference dynamics for $t\le 0$, which takes the form:
\begin{subequations}
\label{eq:Kr0}
\begin{align}
d q(t) &= \frac{p}{m}\, dt , \label{eq:Kr0-1} \\
d p(t) &= - m\omega_0^2 q \, d t - \gamma~ p\, dt + \sqrt{2\gamma m \beta_0^{-1}}\, d W(t) . \label{eq:Kr0-2}
\end{align}
\end{subequations}
At time $t=0^+$ the joint perturbation $\omega_0\rightarrow \omega$ and $\beta_0\rightarrow \beta$ makes the reference dynamics turn into the perturbed one, defined for $t>0$ as follows:
\begin{subequations}
\label{eq:Kr1}
\begin{align}
d q(t) &= \frac{p}{m}\, dt , \label{eq:Kr1-1} \\
d p(t) &= - m\omega^2 q \, d t - \gamma~ p\, dt + \sqrt{2\gamma m \beta^{-1}}\, d W(t) . \label{eq:Kr1-2}
\end{align}
\end{subequations}
The Fokker-Planck equation describing, for $t>0$, the evolution of the probability density $f_t(q,p)$ (an instance of the Kramers equation \cite{Bala}) reads
\begin{equation}
    \frac{\partial f_t}{\partial t}= -\frac{p}{m} \frac{\partial f_t}{\partial q}+m\omega_0^2q\frac{\partial f_t}{\partial p}+\gamma f_t+\gamma p\frac{\partial f_t}{\partial p}+\frac{\gamma m}{\beta}\frac{\partial^2 f_t}{\partial p^2} ,
    \label{FP3}
\end{equation}
and is supplied with the initial datum 
\begin{equation}
    f_0(q,p)=\frac{\beta_0 \omega_0}{2\pi}e^{-\frac{\beta_0}{2} \left(m\omega_0^2 q^2+\frac{p^2}{m}\right)} .
    \label{f0un}
\end{equation}
The dissipation function $\Omega_{\omega,\beta}^{f_0}$ derived from \eqref{FP2} corresponds to the state function
\begin{equation}
   \Omega_{\omega,\beta}^{f_0}(q,p)=A_1 p^2 + B_1 q p + C_1 ,
   \label{Omunder}
\end{equation}
where $A_1 = -\frac{\gamma \beta_0}{m} \left(1 - \frac{\beta_0}{\beta}\right)$, $B_1= \beta_0 (\omega_0^2 - \omega^2)$ and $C_1=-A_1 \langle p^2\rangle_0$, with $\langle p^2 \rangle_0=m/\beta_0$.
We note that $f_0(q,p)$ in \eqref{f0un} is invariant for the reference dynamics \eqref{eq:Kr0}, for $\Omega_{\omega_0,\beta_0}^{f_0}=0$ holds.
Furthermore, letting  \( x = (q, p)^T \in \mathbb{R}^2 \), Eq. \eqref{FP3} can be rewritten in the form
\begin{equation}
    \frac{\partial f_t}{\partial t}=\sum_{i,j=1}^2\bigg[-A_{i,j}\frac{\partial}{\partial x_i}(x_j f_t)+D_{i,j}\frac{\partial^2 f_t}{\partial x_i \partial x_j}\bigg],
\end{equation}
where the matrices $A$ and $D$ have the following form:
\begin{equation}
        A=\begin{pmatrix}
        0 && 1/m \\
        -m\omega^2 && -\gamma 
            \end{pmatrix}
            \quad , \quad
         D=\begin{pmatrix}
        0 && 0 \\
        0 && \gamma/\beta
    \end{pmatrix} .
    \label{matrixAD}
\end{equation}
In the large friction regime $\gamma>2\omega$ that is considered here, the two eigenvalues of the matrix $A$ are real, $-\lambda_{\pm}<0$, with $\lambda_{\pm}=(\gamma\pm\sqrt{\gamma^2-4 \omega^2})/2$. The time scale separation, in this model, is quantified by $\Delta=\lambda_{+}-\lambda_{-}=\sqrt{\gamma^2-4\omega^2}$.

Proceeding as in Sec.~\ref{sec:sec4}, let us consider, first, the exact response formula for the observable $O(q,p)\equiv q^2$.

The evaluation of time correlation functions for the underdamped dynamics \eqref{eq:Kr1} $\langle O(x_s) \Omega_{\omega\beta}^{f_0}(x_0)\rangle_0$ can be carried out by knowledge of the conditional probability density $p(y,t|x,0)$, which in this case reads
\begin{equation}
p(y,t|x,0) = \frac{1}{2\pi\sqrt{\det\Sigma(t)}} \exp\left(-\frac{1}{2}\left(y-e^{A t}x \right)\Sigma^{-1}(t)\left(y-e^{A t}x \right)  \right),
   \label{condprob2}
\end{equation}
where
\[
      \Sigma(t)=\int_0^t e^{A s}\ D\ e^{A^T s} ds
\]
is the covariance matrix. We point out that the stationary covariance matrix $\bar{\Sigma}:=\lim_{t\rightarrow \infty} \Sigma(t)$ satisfies the Lyapunov equation.
\[
A\bar{\Sigma}+\bar{\Sigma}A^T=-D,
\]
which is an instance of the Fluctuation-Dissipation relation \cite{Risken,Pavl,colan25}.
Using Eq.~\eqref{eq:chandra5} in Appendix~\ref{app:appB}, we find that, for the state function \( O(q,p) \equiv q^2 \), formula~\eqref{final} evaluates to:

\begin{eqnarray}
\langle q^2 \rangle_t &=&  \frac{1}{\beta_0 m \omega_0^2} + \frac{e^{-2 t (\lambda_- + \lambda_+)}}{ m \beta \beta_0 \lambda_- (\lambda_- - \lambda_+)^2 \lambda_+ (\lambda_- + \lambda_+) \omega_0^2}\times\nonumber\\
&&    \left[
        e^{2 t \lambda_-}
        \lambda_- (\lambda_- + \lambda_+)
        \left(
            \beta \lambda_- \omega^2 - 
            (\beta_0 \gamma + \beta (-\gamma + \lambda_-)) \omega_0^2
        \right)
        \right. \nonumber\\
     &+& e^{2 t \lambda_+}
        \lambda_+ (\lambda_- + \lambda_+)
        \left(
            \beta \lambda_+ \omega^2 - 
            (\beta_0 \gamma + \beta (-\gamma + \lambda_+)) \omega_0^2
        \right) \nonumber
    \\ &-& 2 e^{t (\lambda_- + \lambda_+)}
        \lambda_- \lambda_+
        \left(
            \beta (\lambda_- + \lambda_+) \omega^2 - 
            (2 \beta_0 \gamma + \beta (-2 \gamma + \lambda_- + \lambda_+)) \omega_0^2
        \right) \nonumber
    \\ && \left.
        + e^{2 t (\lambda_- + \lambda_+)}
        (\lambda_- - \lambda_+)^2
        \left(
            -\beta (\lambda_- + \lambda_+) \omega^2 + 
            (\beta_0 \gamma + \beta (-\gamma + \lambda_- + \lambda_+)) \omega_0^2
        \right)
    \right] .
    \label{varunder}
\end{eqnarray}
More insight on the formula \eqref{varunder} is gained by inspecting the large-$\gamma$ regime. To this aim we adopt in \eqref{varunder} the scaling previously introduced in Sec.~\ref{sec:sec2}, i.e. $\gamma\rightarrow\gamma/\epsilon$, $t\rightarrow t/\epsilon$. The following asymptotics holds
\begin{equation}
\lambda_+ = \frac{\gamma}{\epsilon} + O(\epsilon) \quad , \quad \lambda_-=\frac{\omega^2}{\gamma}\epsilon+o(\epsilon) 
\label{eigenval}.
\end{equation}
Note that the time scale separation \( \Delta \sim \epsilon^{-1} \) diverges as \( \epsilon \rightarrow 0 \), which corresponds to the regime of perfect time scale separation.
Inserting the expressions \eqref{eigenval} in \eqref{varunder} and then taking the overdamped limit $\epsilon\rightarrow 0$, one obtains
\begin{equation}
    \langle q^2\rangle_t= \frac{\beta \omega^2 e^{-2\frac{\omega^2}{\gamma} t}+\left(1-e^{-2\frac{\omega^2}{\gamma} t}\right)\beta_0 \omega_0^2}{m \beta_0 \beta \omega_0^2 \omega^2} \label{varunder2} ,
\end{equation}
which recovers formula \eqref{varover}.
This result highlights a key feature of the response theory developed in Sec.~\ref{sec:sec2}: for certain observables \( O(q,p) \), the two operations of taking the  limit \( \epsilon \rightarrow 0 \) and deriving the response formula~\eqref{final} commute.

We now turn our attention to the state function \( O(q,p) = p^2 \). In this context, the momentum \( p(t) \) is regarded as a \emph{fast} variable, while the rescaling \( t \rightarrow t/\epsilon \) corresponds to observing the evolution of the system on a \emph{slow} time scale~\cite{Jones1995}. On such a scale, as \( \epsilon \) decreases, the quantity \( \langle p^2 \rangle_t \) is expected to thermalize increasingly rapidly toward its equilibrium value \( m/\beta \), as dictated by the equipartition theorem. This behavior is fully confirmed by our calculations, which, using Eq.~\eqref{eq:chandra4}, show that

\begin{eqnarray}
    \langle p^2 \rangle_t &=& \frac{1}{\beta_0 m} 
\Bigg(
1 
+ \frac{
    e^{-2 t (\lambda_- + \lambda_+)} 
    (\beta - \beta_0) \gamma 
}{
    \beta 
    (\lambda_- - \lambda_+)^2 
    (\lambda_- + \lambda_+)
} \Big[
        -e^{2 t (\lambda_- + \lambda_+)} 
        (\lambda_- - \lambda_+)^2 \nonumber\\
        &-& 4 e^{t (\lambda_- + \lambda_+)} 
        \lambda_- \lambda_+ 
        + (\lambda_- + \lambda_+) 
        \big(
            e^{2 t \lambda_+} \lambda_- 
            + e^{2 t \lambda_-} \lambda_+
        \big)
    \Big] \nonumber\\
&+& \frac{
    e^{-2 t (\lambda_- + \lambda_+)} 
    (e^{t \lambda_-} - e^{t \lambda_+})^2 
    \omega^2 
    (\omega - \omega_0)
    (\omega + \omega_0)
}{
    (\lambda_- - \lambda_+)^2 
    \omega_0^2
}
\Bigg) .
\label{p2av}
\end{eqnarray}
At $t=0$ formula \eqref{p2av} correctly reproduces $\langle p^2 \rangle_0=m/\beta_0$.
Letting $\gamma\rightarrow\gamma/\epsilon$, $t\rightarrow t/\epsilon$, using \eqref{eigenval}, and finally passing to the limit $\epsilon\rightarrow 0$, we find
\begin{equation}
 \langle p^2 \rangle_t = \frac{m}{\beta}
 \label{p2avb} \quad , \quad t>0,
\end{equation}
which confirms that, after the rescaling, the quantity $\langle p^2 \rangle_t$ undergoes an instantaneous thermalization. 

As a next step, we consider the state function $O(q,p)=q p$. Using Eq.~\eqref{eq:chandra3} in App. \ref{app:appB}, we find:
\begin{eqnarray}
    \langle q p \rangle_t &=& \frac{e^{-2 t (\lambda_- + \lambda_+)}}{2 \beta \beta_0 \lambda_- (\lambda_- - \lambda_+)^2 \lambda_+ (\lambda_- + \lambda_+) \omega_0^2}\times \nonumber\\
&&\left[
-e^{2 t (\lambda_- + \lambda_+)} \beta (\lambda_- - \lambda_+)^2 (\lambda_- \lambda_+ - \omega^2) (\omega^2 - \omega_0^2) \right. \nonumber\\
&-& e^{2 t \lambda_+} \lambda_+ (\lambda_- + \lambda_+) \bigl[\beta \omega^2 (\lambda_- \lambda_+ + \omega^2)  \nonumber \\
&-& (2 \beta_0 \gamma \lambda_- + \beta (-2 \gamma \lambda_- + \lambda_- \lambda_+ + \omega^2)) \omega_0^2 \bigr] \nonumber\\
&-& e^{2 t \lambda_-} \lambda_- (\lambda_- + \lambda_+) \bigl[\beta \omega^2 (\lambda_- \lambda_+ + \omega^2) \nonumber \\
&-& (2 \beta_0 \gamma \lambda_+ + \beta (-2 \gamma \lambda_+ + \lambda_- \lambda_+ + \omega^2)) \omega_0^2 \bigr] \nonumber\\
&+& 2 e^{t (\lambda_- + \lambda_+)} \lambda_- \lambda_+ \bigl[\beta \omega^2 (\lambda_-^2 + \lambda_+^2 + 2 \omega^2) \nonumber \\
&-& \left. (2 \beta_0 \gamma (\lambda_- + \lambda_+) + \beta (\lambda_-^2 + \lambda_+^2 
- 2 \gamma (\lambda_- + \lambda_+) + 2 \omega^2)) \omega_0^2 \bigr]
\right] ,
\end{eqnarray}
which, after rescaling $\gamma\rightarrow\gamma/\epsilon$, $t\rightarrow t/\epsilon$, and passing to the limit $\epsilon\rightarrow 0$, yields
\begin{equation}
    \langle q p \rangle_t = 0 \quad , \quad t>0.
\end{equation}
Finally, we consider the state function $O(q,p)\equiv\Omega_{\omega,\beta}^{f_0}(q,p)$.
From Eq.~\eqref{Omunder} we can write
\begin{equation}
    \langle \Omega_{\omega,\beta}^{f_0}\rangle_t = A_1 \langle p^2 \rangle_t + B_1 \langle q p \rangle_t + C_1.
\end{equation}
At $t=0$ one correctly finds
\[
\langle \Omega_{\omega,\beta}^{f_0}\rangle_t = A_1 \langle p^2 \rangle_0+C_1=0 .
\]
Introducing the usual scaling $\gamma\rightarrow\gamma/\epsilon$, $t\rightarrow t/\epsilon$, in the limit \( \epsilon \rightarrow 0 \), we obtain
\begin{eqnarray}
\langle\Omega_{\omega,\beta}^{f_0}\rangle_t&=&A_1 \langle p^2 \rangle_t+C_1\quad , \quad t>0 \nonumber\\
&=&\frac{\gamma}{\beta^2}(\beta-\beta_0)^2 .
     \label{Omunder2}
\end{eqnarray}
We observe that the quantity \( \langle \Omega_{\omega,\beta}^{f_0} \rangle_t \) undergoes an instantaneous thermalization in the appropriate scaling limit: it exhibits a discontinuous jump from the value \( 0 \) at \( t = 0 \) to the final value given by Eq.~\eqref{Omunder2}.

We further note that Eq.~\eqref{Omunder2} does not recover the expression in Eq.~\eqref{resp}. This discrepancy arises because the observables defined in Eqs.~\eqref{Omover} and~\eqref{Omunder}, although both referred to as Dissipation Functions, correspond to distinct state functions. 

\section{Conclusions}
\label{sec:sec6}

In this work, we have extended the response theory based on Dissipation Functions, originally developed in the context of deterministic dissipative dynamical systems~\cite{Jepps16}, to stochastic processes with additive noise governed by a Fokker--Planck equation. A key feature of our derivation is twofold: (i) the response formulae are conveniently expressed in terms of the initial probability density \cite{Jepps16}; and (ii) these formulae remain valid for perturbations of arbitrary magnitude, and are therefore not restricted to the conventional linear response regime.

Focusing on the analytically tractable case of particles confined in a harmonic potential, we have demonstrated a notable property of the formalism: for specific observables, the operations of deriving the response formulae and taking the limit \( \epsilon \to 0 \), known as the overdamped limit, commute. Moreover, the theory correctly captures the average behavior of certain state functions, such as the Dissipation Function associated with the underdamped dynamics, which exhibits instantaneous relaxation to its equilibrium value in the overdamped limit; see Eq.~\eqref{Omunder2}. While this behavior is consistent with model reduction procedures such as adiabatic elimination, which rely on a large separation of time scales~\cite{Gardiner09,Sekimoto10}, it does not reproduce the behavior of the average Dissipation Function obtained for the overdamped dynamics; see Eq.~\eqref{resp}. This discrepancy arises from the fact that the two Dissipation Functions depend on different sets of variables, namely fast variables in the underdamped case and slow variables in the overdamped dynamics, see Eqs.~\eqref{Omunder} and~\eqref{Omover}, respectively. As a result, their averages exhibit distinct behavior in the overdamped limit.

Although such discrepancies are often encountered in the study of anomalous transport phenomena~\cite{LLP03}, they can more fundamentally be traced back to the distinct time scales governing the decay of correlation functions, as explicitly derived for the Brownian oscillator model in Secs.~\ref{sec:sec4} and~\ref{sec:sec5}.

Looking ahead, the framework developed in this work sets the stage for applying response theory to high-dimensional stochastic systems, where model reduction techniques may assist in identifying a relevant subset of resolved variables, typically those evolving on slow time scales and governing the macroscopic behavior of the system.

\appendix
\setcounter{equation}{0}

\section[\appendixname~\thesection]{Derivation of the Dissipation Function for Fokker-Planck  dynamics}
\label{app:appA}
We sketch the derivation of the Dissipation Function $\Omega^{f_t}(x):\mathbb{R}^d\rightarrow \mathbb{R}$, defined as: 
\[
\Omega^{f_t}(x) = \frac{L^* f_t(x)}{f_t(x)} \; ,
\] 
expressed in terms of the Fokker-Planck operator $L^*$ in $\mathbb{R}^d$, i.e.:
\begin{equation}
  L^* f_t(x) = \nabla \cdot \left[-A(x)f_t(x) + \nabla \cdot (D(x)f_t(x)) \right] \,,  
  \label{FPop}
\end{equation}
cf. Eq. \eqref{forwop}.
Using the identities:
\begin{align*}
\nabla \cdot (A(x)f_t(x)) &= (\nabla \cdot A(x)) f_t(x) + A(x) \cdot \nabla f_t(x) \\
\nabla \cdot \left( \nabla \cdot (D(x)f_t(x)) \right) &= \nabla \nabla : (D(x)f_t(x)) \; ,
\end{align*}
we obtain
\[
L^*f = -(\nabla \cdot A(x)) f_t(x) - A(x) \cdot \nabla f_t(x) + \nabla \nabla : (D(x)f_t(x)) \; ,
\]
which yields
\[
\Omega^{f_t}(x) = -\nabla \cdot A(x) - A(x) \cdot \nabla \log f_t(x) + \frac{1}{f_t(x)} \nabla \nabla : (D(x)f_t(x)) \; .
\]
Then, expanding the last term
\[
\frac{1}{f_t(x)} \nabla \nabla : (D(x)f_t(x)) 
= \nabla \nabla : D(x) 
+ 2 \frac{(\nabla \cdot D(x)) \cdot \nabla f_t(x)}{f_t(x)} 
+ \frac{D(x) : \nabla \nabla f_t(x)}{f_t(x)}
\]
and using the identity
\[
\frac{\nabla f_t(x)}{f_t(x)} = \nabla \log f_t(x), 
\quad 
\frac{\nabla \nabla f_t(x)}{f_t(x)} = \nabla \nabla \log f_t(x) + \nabla \log f_t(x) \otimes \nabla \log f_t(x)
\]
we obtain
\begin{eqnarray}
\Omega^{f_t}(x) 
&= -\nabla \cdot A(x) 
- A(x) \cdot \nabla \log f_t(x) 
+ \nabla \nabla : D(x) \nonumber\\
&\quad + 2\, (\nabla \cdot D(x)) \cdot \nabla \log f_t(x) \nonumber\\
&\quad + D(x) : \left[ \nabla \nabla \log f_t(x) + \nabla \log f_t(x) \otimes \nabla \log f_t(x) \right] \; ,
\label{Om2}
\end{eqnarray}
which recovers Eq. \eqref{Om}
In dimension $d=1$ Eq. \eqref{Om2} reduces to
\begin{equation}
\Omega^{f_t}(x) = -A'(x) - A(x) \frac{f_t'(x)}{f_t(x)} + D''(x) + 2 D'(x) \frac{f_t'(x)}{f_t(x)} + D(x) \frac{f_t''(x)}{f_t(x)}  \; .
    \label{Om3}
\end{equation}

Furthermore, in the presence of constant $A(x)=A$ and $D(x)=D$, Eq. \eqref{Om2} yields
\begin{equation}
\Omega^{p_t}(x) = - A \cdot \nabla \log f_t(x) 
+ D : \nabla \nabla \log f_t(x) 
+ D : \left( \nabla \log f_t(x) \otimes \nabla \log f_t(x) \right) \; .
    \label{Om4}
\end{equation}
which admits an even further simplification for $d=1$:
\begin{equation}
   \Omega^{f_t}(x) = -A \frac{f_t'(x)}{f_t(x)} + D \frac{f_t''(x)}{f_t(x)} \; .
    \label{Om5}
\end{equation}

\section[\appendixname~\thesection]{Conditional averages for the underdamped dynamics}
\label{app:appB}

Using the setup introduced in Sec.~\ref{sec:sec5}, let \( O:\mathbb{R}^2 \to \mathbb{R} \) be a generic state function, and define
\[
\overline{O}(x,t) = \int O(y)\, p(y,t \mid x,0)\, dy
\]
as its conditional expectation, computed with respect to the conditional probability density \( p(y,t \mid x,0) \), as given in Eq.~\eqref{condprob2}.
From the analysis of Chandrasekhar~\cite{chandra}, in the regime $\gamma > 2 \omega$ one obtains the following results:

\begin{subequations}
\label{eq:chandra}
\begin{align}
\overline{q_t} &= e^{-\frac{\gamma t}{2}} 
    \left( \cosh\left(\frac{\Delta t}{2}\right) 
    + \frac{\gamma}{\Delta} 
    \sinh\left(\frac{\Delta t}{2}\right) \right) q + \frac{2}{\Delta} e^{-\frac{\gamma t}{2}} 
    \sinh\left(\frac{\Delta t}{2}\right) \frac{p}{m} \label{eq:chandra1}\\
\overline{p_t} &= -\frac{2m\omega^2}{\Delta} 
    e^{-\frac{\gamma t}{2}} 
    \sinh\left(\frac{\Delta t}{2}\right) q + e^{-\frac{\gamma t}{2}} 
    \left( \cosh\left(\frac{\Delta t}{2}\right) 
    - \frac{\gamma}{\Delta} 
    \sinh\left(\frac{\Delta t}{2}\right) \right) p \label{eq:chandra2}\\
\overline{q_t p_t} &= \overline{q_t} \  \overline{p_t} + \frac{4}{\beta} \frac{\gamma}{\Delta^2} e^{-\gamma t} 
    \sinh^2\left(\frac{\Delta t}{2}\right) \label{eq:chandra3}\\
\overline{p_t^2} &= \overline{p_t}^2 + \frac{m}{\beta} \left( 1 - e^{-\gamma t} 
    \left( 2\left(\frac{\gamma}{\Delta}\right)^2 
    \sinh^2\left(\frac{\Delta t}{2}\right) 
    - \frac{\gamma}{\Delta} 
    \sinh\left(\Delta t\right) + 1 \right) \right) \label{eq:chandra4}\\
\overline{q_t^2} &= \overline{q_t}^2 + \frac{1}{m\beta\omega^2} \left( 1 - e^{-\gamma t} 
    \left( 2\left(\frac{\gamma}{\Delta}\right)^2 
    \sinh^2\left(\frac{\Delta t}{2}\right) 
    + \frac{\gamma}{\Delta} 
    \sinh\left(\Delta t\right) + 1 \right) \right) \label{eq:chandra5} \; ,
\end{align}
\end{subequations}
with $\Delta=\lambda_+-\lambda_-$.

\bigskip

\textbf{Acknowledgements} This work was carried out under the auspices of the Italian National Group of Mathematical Physics. The research of MC has been developed in the framework of the research project National Centre for HPC, Big Data and Quantum Computing - PNRR Project, funded by the European Union - Next Generation EU.

 \bibliography{biblio}

\end{document}